\documentclass[%
 reprint, 
 amsmath,amssymb,
 aps,
]{revtex4-1}
\usepackage{graphicx}
\usepackage{listings}
\usepackage{dcolumn}
\usepackage{bm}
\usepackage{xcolor}
\usepackage{upgreek}

\usepackage{url}

\begin{document}


\title{Power-law decay of force on cell membrane tethers reflects long-ranged relaxation of membrane tension}

\author{Emeline Laborie}
 \affiliation{Laboratoire de Biochimie Th\'eorique (LBT), Universit\'e Paris Cit\'e \& CNRS (UPR 9080),
 	Paris, France}
 \affiliation{Institute of Applied Physics, TU Wien, A-1040 Wien, Austria}
 	%
\author{Andrew Callan-Jones}%
 \email{\href{andrew.callan-jones@u-paris.fr}{andrew.callan-jones@u-paris.fr}}
\affiliation{%
	Laboratoire Mati\`ere et Syst\`emes Complexes (MSC), Universit\'e Paris Cit\'e \& CNRS (UMR 7057), Paris, France
}%
\date{\today}

\begin{abstract}\vspace{0.1cm}
	\begin{center}
		\Large{Abstract}
	\end{center}

	 Membrane tension is an acknowledged regulator of a wide array of cell functions; however, whether it acts locally or globally in different contexts remains debated. Recent experiments that locally perturb tension and measure the response elsewhere on the membrane are not conclusive, and the mechanism of diffusive tension relaxation has been called into doubt.  Here, we consider the tension response to a sudden extension of a membrane tether, and report a quantitative signature of dynamic tension relaxation, which up to now is missing. 
	 We present a theory based on tension diffusion leading to a prediction of power-law decay in time of the force holding the tether, with a material-independent exponent of $1/3$. This prediction is confirmed to within a few percent by re-analyzing eleven sets of tether data from two cell types with distinct membrane cortical architectures.  Overall, our scaling results indicate the absence of a relevant characteristic length and therefore generically reflect tension relaxation by long-range membrane lipid flows.
\end{abstract}

\maketitle


Membrane tension influences a wide range of biological functions at different scales, including intracellular trafficking, differentiation~\cite{Li:2023uz}, cell motility~\cite{Sens:2015vu,Garcia-Arcos:2024tf}, cancer metastasis~\cite{Itoh:2023aa}, and wound healing~\cite{raj2024membrane}.  
How fast changes in tension propagate is now hotly debated, and bears upon whether tension is a local or global regulator of cell function~\cite{SITARSKA202011,Ghisleni:2024wv} 

The question of \emph{how} local changes in tension propagate was first taken up in Shi and co-workers~\cite{Shi:2018tv}.  Their work built upon the long-known causal link between gradients in membrane tension and membrane flows~\cite{Dai:1995wr}.  They described the lipid membrane studded with immobile transmembrane proteins linked to an adjacent cortex, providing frictional resistance,  as a two-dimensional porous medium, and the membrane as an isotropically stretchable surface, and showed, in theory, 
tension inhomogeneities should relax diffusively. 
To test this, Shi et al. used a double-tether assay~\cite{Dai:1995wr}, whereby the plasma membrane tension changes caused by perturbations to one tether could, in principle, be detected by a second one a fixed distance away.  With this method, studies aimed at measuring speed of tension propagation using double tethers on different cell types and conditions have not been conclusive.  Slow, or no propgation, has been observed on HeLa cells, fibroblasts, MDCK cells~\cite{Shi:2018tv}; on neutrophils~\cite{De-Belly:2023vo}; on chromaffin cells~\cite{Gomis-Perez:2022vx}; on HEK cells under isotonic conditions~\cite{Dharan:2025tx}; and on neural dendrites~\cite{Shi:2018tv,Shi:2022uj}.  In contrast, rapid propagation has been observed on axons~\cite{Shi:2022uj,catala2025measuring}; on HEK cells under hyptonic conditions~\cite{Dharan:2025tx}; on bipolar cell terminals~\cite{Gomis-Perez:2022vx}; and from tension changes caused by actin-based protrusions of the plasma membrane~\cite{De-Belly:2023vo}. 
To explain these binary findings, there
 have been recent results from simulations~\cite{Shi:2022uj,Gomis-Perez:2022vx,catala2025measuring,Paraschiv:2021ub} and theory~\cite{Dharan:2025tx,catala2025measuring} whose predictions have been tested qualitatively with the double-tether assay.
 In all cases, however, no \emph{quantitative} signature of diffusive propagation has been found, and the diffusion mechanism itself has been called into question~\cite{Gomis-Perez:2022vx}.  
%


\begin{figure*}
	\centering
	\includegraphics[width=.9\textwidth]{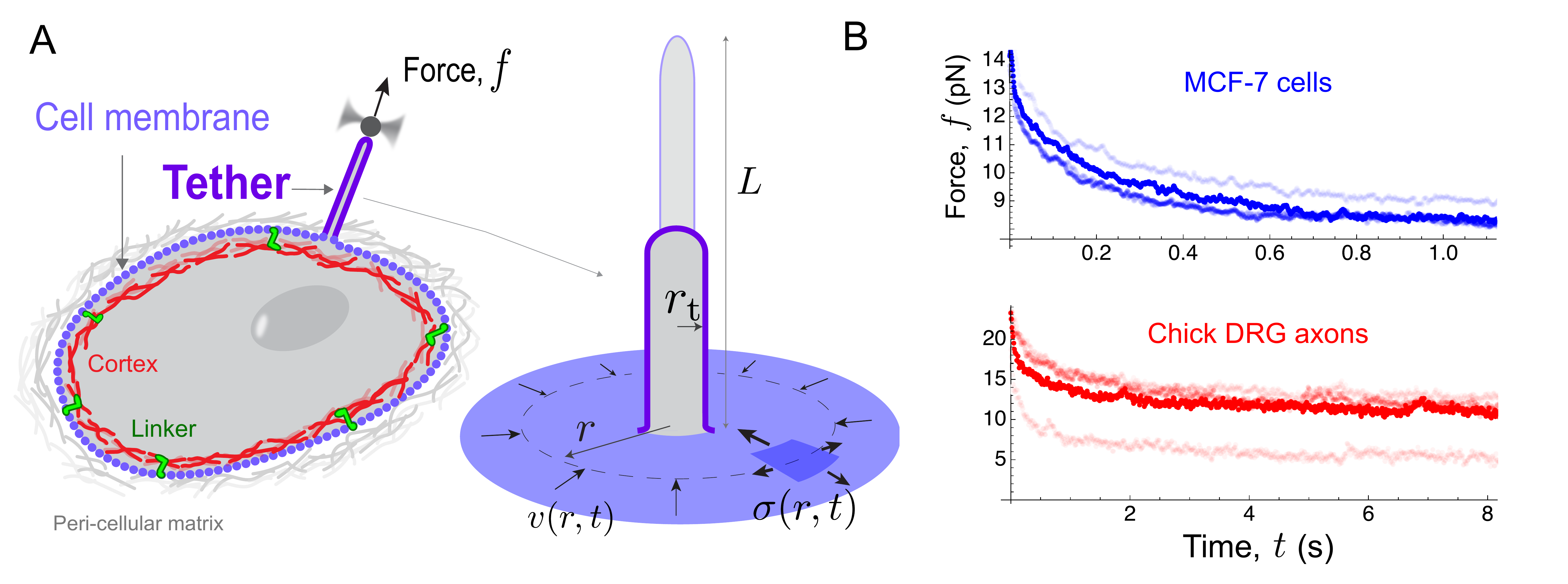}
	\caption{{\bf Dynamic pulling to probe cell membrane tension. } A: Illustration of a cell membrane tether, held by a force $f$, and suddenly extended to a new length, $L$. According to our model, friction between membrane and cell cortex and/or peri-cellular matrix causes an abrupt initial jump in tension ($\sigma$), at radial position $r=r_{\rm t}$, which relaxes by plasma membrane flows, $v(r,t)$. 
		B:  Force relaxation versus time, $t$, after sudden extension of a tether, at time $t=0$, from the plasma membrane of MCF-7 breast cancer cells (upper, blue) and from chick axons (lower, red).  Each trace is from a different tether extension experiment. MCF-7 and axon data are courtesy of Pradhan et al.~\cite{Pradhan:2022tf} and Datar et al.~\cite{Datar:2015vp}.}
	\label{fig:Fig1}
\end{figure*}

To find this signature, here we revisit a simpler, well-known, dynamic tension assay, in which a pre-existing, static tether is suddenly extended from a length $L-\Delta L$ to a new one, $L$, which is then held constant.
The corresponding force data show a common feature: a fast initial decay followed by a much slower one.  Previous attempts to explain this have been empirical or phenomenological~\cite{Li:2002vn,Heinrich:2005vs,Jauffred:2007uj,
Campillo:2013ur,Gomis-Perez:2022vx,Pradhan:2022tf,Li:21Optics}.  A typical approach has been to fit a sum of two exponential functions, with a small and large decay time, respectively.  There is, however, no theoretical basis for this.  A different attempt to model observed force decay has been to assume tension relaxes by in-plane osmotic equilibration of mobile membrane proteins between tether and surrounding membrane~\cite{Datar:2015vp,Paraschiv:2021ub}.  This assumes tension, however, is instantaneously equilibrated on the membrane, which is difficult to reconcile with the recognized frictional resistance felt by the surrounding membrane~\cite{Shi:2018tv}.  Significantly, these models do not account for long-range membrane flows associated with tension gradients, which are obstructed by the cytoskeleton, or more generally from membrane-adhered structures such as extra/pericellular matrix or a neighboring cell~\cite{Ghisleni:2024wv}.  Given the acknowledged role of membrane flows in transmitting tension, we think there is a strong need for a theory with readily testable predictions (using an optical tweezer set-up, for example) that can relate tension gradients sustained by membrane-adjacent structures to an observable, namely the tether force.

We thus set out to find the physical origin and mathematical nature of the decay of force on a rapidly extended tether. 
We posed the problem of a suddenly extended tether as a diffusion boundary value problem with an initial tension ``shock''. 
This problem is difficult because of the moving boundary, namely the tether radius, which evolves in time.  We used numerical and analytical approaches to solve it.  Our solution leads to a testable quantitative prediction of the time-dependent force holding the tether.
 We find it decays as a power law in time with exponent $\delta=1/3$, which is independent of material properties.  This prediction is confirmed remarkably well by fitting our model to 11 sets of existing force data from two cell types with distinct cortical architectures~\cite{Datar:2015vp,Pradhan:2022tf}.

%

\section{Model and Results}
\label{sec:ModelandResults}
%
%
The experimental set-up motivating our model is depicted in Fig.~\ref{fig:Fig1}A. A cylindrical tether, with initial radius $r_0$, pulled from the cell plasma membrane, initially at uniform tension $\sigma_0$, is rapidly extended from an initial length $L_0$ to a new one $L>L_0$ at a time $t=0$, and then held at $L$ for $t>0$.  
Frictional
resistance prevents immediate membrane response, and the tether extends initially at roughly constant area.  Furthermore, the tether radius $r_{\rm t}(t) $ adjusts rapidly to the tension $\sigma_{\rm t}(t)$ according to the mechanical relation
\begin{equation}
r_{\rm t}=\sqrt{\kappa/2\sigma_{\rm t}}\,,
\label{eq:TetherRadius}
\end{equation}
where $\kappa$ is the membrane bending modulus~\cite{Brochard-Wyart:2006wv}.  We note $r_0$ and $\sigma_0$ are related via the above equation for $t<0$. 
From this equation, sudden tether extension at constant area means its tension increases rapidly and is given in terms of the extension factor $\lambda = L/L_0$ by
\begin{equation}
\sigma_{\rm t}(0) = \lambda^2\sigma_0\,.
\label{eq:sigmajumpinitial}
\end{equation}
A second mechanical relation gives the force on the tether in terms of its radius~\cite{Brochard-Wyart:2006wv}:
\begin{equation}
f=2\pi\kappa/
r_{\rm t}\,.
\label{eq:Force}
\end{equation}
%
Because of the large tension gradient at the tether position at early times, lipids from the surrounding membrane flow into the tether, $r_{\rm t}$ increases, and the force decays.  Tether pulling experiments on the plasma membrane of MCF-7 breast cancer cells~\cite{Pradhan:2022tf} and of the axons of chick dorsal root ganglion neurons~\cite{Datar:2015vp}, for example, show typical force decay (Fig.~\ref{fig:Fig1}B).  These data show a rapid decay at early times followed by a slower one at long times. The data  curves for the two cell types show a similar form, despite the experimental time scales being different. 
\begin{figure*}
	\centering
	\includegraphics[width=.9\textwidth]{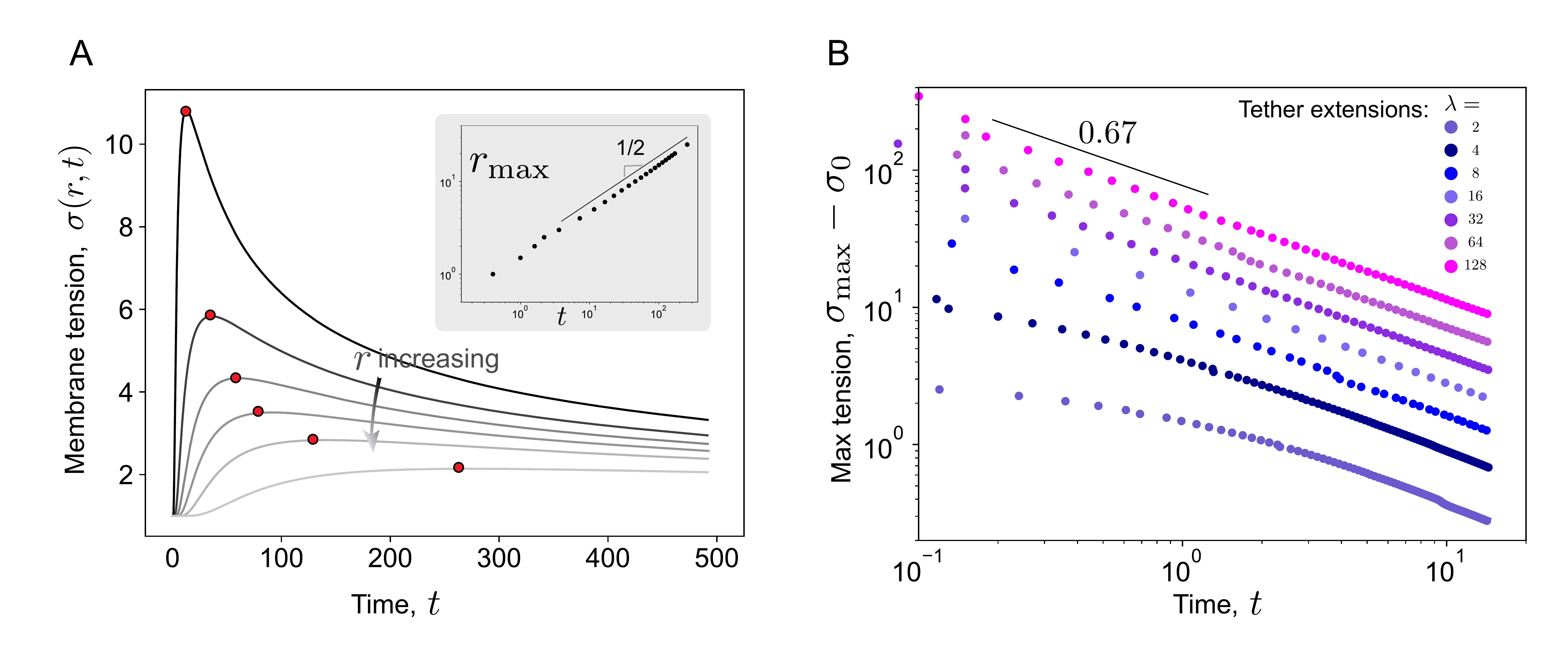}
	\caption{{\bf Membrane tether as a source of tension diffusion}.  A: Membrane tension versus $t$, at different radial positions, $r$, from the tether center.  The red points indicate the maximum tension in time.  Inset: the position of the maximum tension, $r_{\rm max}$, versus $t$ scales as $r_{\rm max}\sim t^{1/2}$, typical of a diffusion problem.    
		B: The value of maximum tension in time versus $t$.  With increasing tether extension, $\lambda$, a scaling $\sigma_{\rm max}\sim t^{-0.67}$ is numerically observed.  Parameter values used to obtain this figure are given in Materials and Methods. 
	}
	\label{fig:Fig2}
\end{figure*}

In our model, the flow of lipids at a position $r$ from the tether (Fig.~\ref{fig:Fig1}A) is described by a velocity field $\mathbf{v}$ proportional to the tension gradient there:
$\mathbf{v}=\mu\nabla\sigma$, where $\mu$ is a constant coefficient describing the mobility of lipids, and is inversely proportional to the membrane friction with its surroundings.  The symmetry of the problem dictates $\mathbf{v}$ is centripetal and the tension depends spatially only on the radial coordinate as measured from the tether: $\sigma=\sigma(r,t)$.  The local rate of compression (or expansion) of the membrane 
is given by minus the divergence of the velocity, denoted $\nabla\cdot\mathbf{v}$,  assuming there is  no lipid exchange between the membrane and its surroundings.  The resulting relative accumulation (or depletion) of lipid mass lowers (or raises) the tension, with the constant of proportionality being the stretch modulus $E$. As a result, the tension in the tether-surrounding membrane can be shown to satisfy the diffusion equation~\cite{Shi:2018tv}, which is linear in $\sigma(r,t)$:
\begin{equation}
\frac{\partial \sigma}{\partial t} = D \frac{1}{r}\frac{\partial}{\partial r}\left(r\frac{\partial\sigma}{\partial r}\right)\,,
\label{eq:DiffusionEquation}
\end{equation}
where $D = \mu E$ is 
the tension diffusion constant.  

Solving Eq.~\ref{eq:DiffusionEquation} requires specifying boundary conditions. The first assumes the membrane surrounding the tether is large and thus behaves as a lipid reservoir. Therefore, we assume that far away from the tether the tension tends to its original value:
\begin{equation}
\sigma(r\to\infty,t)=\sigma_0\,.
\label{eq:BCfaraway}
\end{equation}
The second condition comes from the fact that, as the extended tether is relaxing, the flux of lipid mass from the membrane at $r=r_{\rm t}(t)$ to the tether must equal the rate of change of its area:
\begin{equation}
2\pi r_{\rm t}\mu\left.\frac{\partial \sigma}{\partial r}\right|_{r=r_{\rm t}(t)} = -2\pi L \frac{d r_{t}}{dt}\,.
\label{eq:BCflux}
\end{equation}
The initial conditions on the tension are discontinuous:  $\sigma(r,t=0)=\sigma_0$ everywhere except at $r_{\rm t}$, where it is given by Eq.~\ref{eq:sigmajumpinitial}.  To fully solve the problem, a last relation is still needed to find the unknown $r_{\rm t}(t)$.  This is given by Eq.~\ref{eq:TetherRadius}, once we recognize 
\begin{equation}
\sigma_{\rm t}(t)=\sigma(r=r_{\rm t}(t),t)\,.
\label{eq:sigmaatrt}
\end{equation}
The boundary condition Eq.~\ref{eq:BCflux} is nonlinear---which can be seen from its left-hand side--- and must be applied at a moving position, $r_{\rm t}(t)$.  This is what makes the problem difficult.  Nonetheless, some insight can be obtained by identifying the relevant dimensionless parameters.  
By expressing lengths in units of $r_0$, times in $r_0^2/D$, and tension in $\sigma_0$, it becomes apparent the two dimensionless parameters governing the problem are the tether extension factor $\lambda$ and a second quantity $b$ defined as
\begin{equation}
 b=\frac{\sqrt{2\kappa \sigma_0}}{E L_0}\,.
\label{eq:bDef}
\end{equation}
%
This quantity emerges from the dimension-free version of Eq.~\ref{eq:BCflux} and controls the rate of change of the tether radius.   
 Using experimental data we can show $b$ is much smaller than one (Materials and Methods), and as a result it is expected the relaxation of $r_{\rm t}$ occurs slowly, and therefore the tether force does as well. 
\begin{figure*}
	\centering
	\includegraphics[width=.9\textwidth]{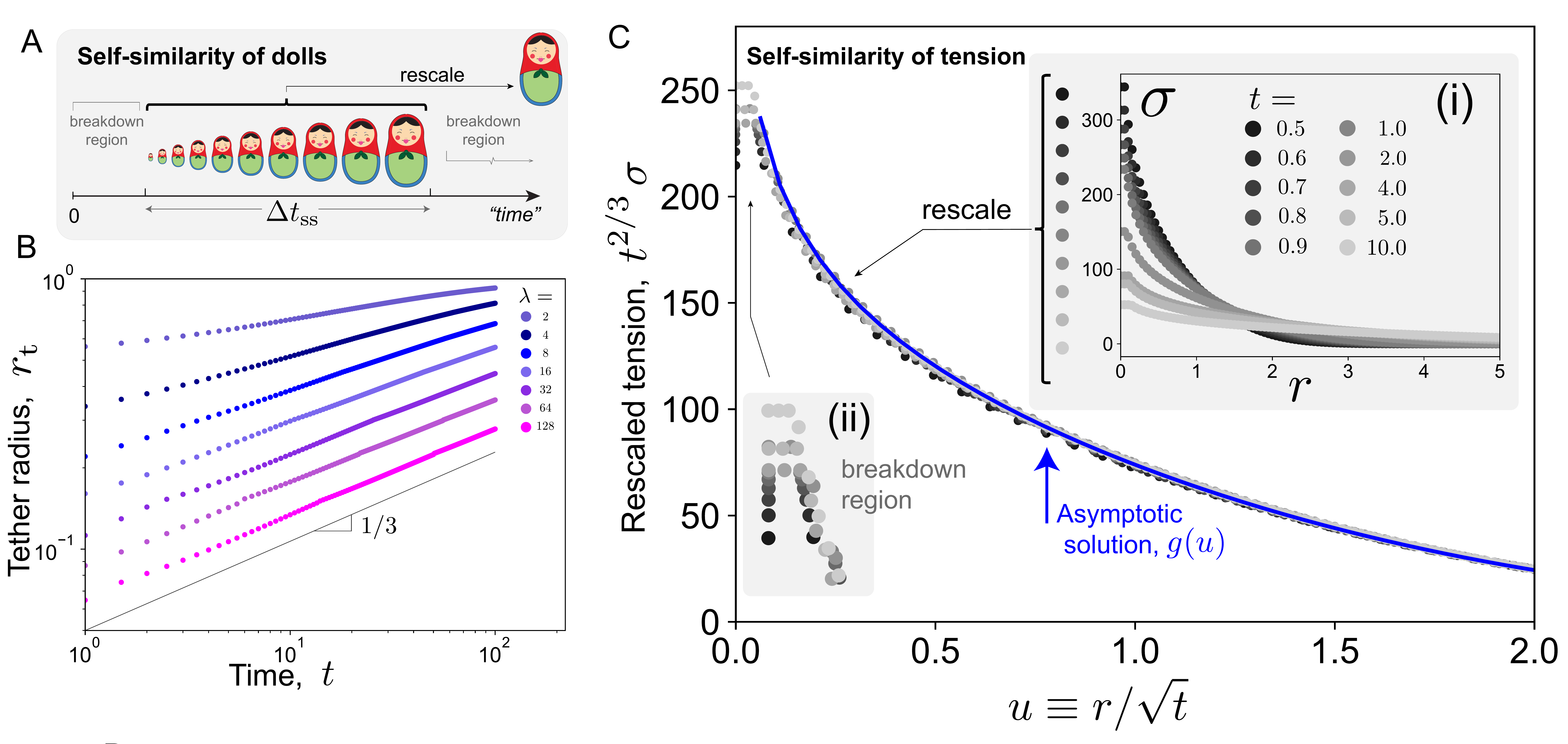}
	\caption{{\bf Model predicts self-similar membrane tension relaxation}. A: Analogy with \emph{Matriochka} of different sizes (``times'').  The dolls are self-similar over a range $\Delta t_{\rm ss}$: by rescaling by an appropriate factor they ``collapse'' onto a single doll.  In real life, self-similarity breaks down outside of $\Delta t_{\rm ss}$, because of fabrications limits, storage limits (at large sizes), etc. B: The tether radius $r_{\rm t}$ versus time, $t$.  With increasing extension $\lambda$ the scaling approaches $r_{\rm t}\sim t^{1/3}$.  C:  The numerical tension versus $r$ at different times $t$ [inset (i)] exhibits self-similarity over a range $\Delta t_{\rm ss}$.  This is seen by collapse of data onto a common curve when plotted in rescaled coordinates, $t^{2/3}\sigma(r,t)$ versus
		$u\equiv r/\sqrt{t}$. The asymptotic expression $g(u)$ [Eqs.~\ref{appendeq:gofuMeijerG} and \ref{appendeq:expressionforA}] is given by blue curve.
		Self-similarity breaks down over a region near the  tether position $r_{\rm t}(t)$ [inset (ii)].  On the diffusion scale this region decreases with time as $t^{1/6}$.
		Parameter values used are given in Materials in Methods.
	}
	\label{fig:Fig3}
\end{figure*}
\subsection{Numerical solution to tension diffusion after sudden tether extension}
\label{subsec:RelaxationbyDiffusion}
We first approached the problem Eqs.~\ref{eq:DiffusionEquation}-\ref{eq:BCflux}
numerically, using an established method for moving boundary problems
~\cite{crank1987free}; see Materials and With this method we determined the membrane tension $\sigma(r,t)$ following a sudden tether extension; see Fig.~\ref{fig:Fig2}A.  These curves resemble those obtained by simulation of tension diffusion~\cite{Shi:2018tv,Gomis-Perez:2022vx}.  Furthermore,  they also resemble classical concentration diffusion from a point source, and since the problem is two-dimensional, it is tempting to assume  $\sigma(r,t)-\sigma_0 \propto t^{-1} e^{-r^2/4 t}$~\cite{ChaikinLubensky}. Indeed, if we follow the position, $r_{\rm max}$, of the maximum tension in time (red points in Fig.~\ref{fig:Fig2}A), we numerically obtain $r_{\rm max}\propto t^{1/2}$, as expected (inset).  However, the value of tension there, $\sigma(r_{\rm max}(t),t)$, does not scale as $t^{-1}$.  Instead, with increasing extension, $\lambda$, we find that 
$\sigma_{\rm max}-\sigma_0$ tends to a power law $\sim t^{-0.67}$; see Fig.~\ref{fig:Fig2}B.  This finding suggests membrane tension relaxation following a sudden tether extension can be characterized as non-classical diffusion due to a tension source.
\subsection{Tension relaxation is self-similar, with testable predictions}
\label{subsec:DiffusionSelfSimilarNonClassical}
To make sense of the previous findings we next looked for an analytical solution to the tension, $\sigma(r,t)$.  This task is hopeless for general values of the dimensionless parameters $\lambda$ and $b$. 
However, the theory of partial differential equations~\cite{barenblatt_1996,lagree2020self} gives us powerful tools to readily obtain a limiting (referred to here as asymptotic) expression which, for increasing $\lambda$ and decreasing $b$, attracts solutions to Eqs.~\ref{eq:DiffusionEquation}-\ref{eq:BCflux}.  The reason is, for large $\lambda$ and over a time window $\Delta t_{\rm ss}$ that excludes very short and long times, the tension near the tether will be much greater than $\sigma_0$. The form of $\sigma(r,t)$ versus $r$ will be insensitive to its (small) value far away, which can be effectively set to zero. Furthermore, for small $b$ the tether tension $\sigma_{\rm t}$ will decay slowly back to $\sigma_0$, and the window $\Delta t_{\rm ss}$ will be long and observable.  As a result, during $\Delta t_{\rm ss}$ the nature of the decay will not depend on $\sigma_0$, which is only felt at very long times.  These observations imply a scale invariance~\cite{barenblatt_1996} of Eqs.~\ref{eq:DiffusionEquation}-\ref{eq:BCflux} and self-similarity of the tension, much like \emph{Matriochka}, as illustrated in Fig.~\ref{fig:Fig3}A.  This property means that $\sigma(r,t)$ will not depend separately on the independent variables $r$ and on $t$, but only on a certain combination of them, thus simplifying the problem greatly.  We thus searched for a self-similar solution for $\sigma(r,t)$ and a power law solution for $r_{\rm t}(t)$:
%
%
%
\begin{align}
\sigma(r,t) &= t^{-\alpha}\,g(r/t^\beta)\quad\textrm{for  $r\ge r_{\rm t}(t)$} \,;
\label{eq:sigmaSelfSimilarAnsatz}\\
r_{\rm t}&=a\,t^{\delta}\,.
\label{eq:rtetherpowerlaw}
\end{align}
In these equations the exponents $\alpha$, $\beta$, and $\delta$ are positive, $g$ is an unknown function of $u\equiv r/t^\beta$, and the coefficient $a$ is an unknown function of $\lambda$ and $b$. 
By substituting Eq.~\ref{eq:sigmaSelfSimilarAnsatz} into the diffusion equation, Eq.~\ref{eq:DiffusionEquation}, first we can  readily show  $\beta=1/2$, as expected for a diffusion problem.
In the next step, detailed in Materials and Methods, we use the boundary conditions Eqs.~\ref{eq:BCflux} and \ref{eq:sigmaatrt} to find $\alpha$ and $\delta$:
\begin{equation}
\alpha=2/3\,,\quad \delta=1/3\,.
\label{eq:exponents}
\end{equation}
Thus, the value of $\alpha$ explains the numerical decay of the maximum tension, $\sigma_{\rm max}$, seen in Fig.~\ref{fig:Fig2}B.  Furthermore, the numerically determined tether radius, with increasing $\lambda$, tends to a power law with slope of 1/3 (Fig.~\ref{fig:Fig3}A), supporting our determination of $\delta$. 

In the last step, with knowledge of $\alpha$, $\beta$, and $\delta$, the quantities $a$ and $g$ can be obtained by solving Eqs.~\ref{eq:DiffusionEquation}-\ref{eq:BCflux}.  The function $g$ is given in terms of $u$, $\lambda$ and $b$ by Eqs.~\ref{appendeq:gofuMeijerG} and \ref{appendeq:expressionforA} in Materials and Methods.  
To test this functional expression, we plotted the numerically determined tension in rescaled coordinates.  From a graph of $t^{2/3}\sigma$ versus $u$ we find that the numerical data collapse onto a common curve, like Russian dolls (Fig.~\ref{fig:Fig3}B).  This collapse is itself a strong validation of the assumption of self-similarity. Furthermore, the analytically determined $g(u)$ (blue curve) falls neatly onto the numerical data, 
with no adjustable parameters. 

Taken together, we have found an asymptotically exact solution for the membrane relaxation after a tether is suddenly extended.  Experimentally, extension factors $\lambda\approx 1-10$ can be achieved and $b$ is much less than one; we expect our limiting calculation to be a robust attractor for a wide range of initial conditions.  
Our model then leads to two quantitatively testable predictions.   First, by measuring the tension in the vicinity of the extended tether, either with secondary probe tethers or with tension-sensitive fluorescent membrane probes~\cite{colom2018fluorescent}, the tension profile could be measured and compared with our self-similar prediction. 
Second, our determination of the exponent $\delta$ and the coefficient $a$ can be directly tested experimentally with existing data, as we next show.


\subsection{Prediction of power law tether force decay is confirmed by experiments
on two cell types}
\label{subsec:PowerLawRelaxation}

The force holding the tether is inversely proportional to the radius, $f\propto 1/r_{\rm t}$, as seen in Sec.~\ref{sec:ModelandResults}.  This means the force on a suddenly extended tether should decay with a power law with exponent 1/3.  Combining this with our determination of $a$ --- see Eq.~\ref{appendeq:FinalEquationfora} of Materials and Methods --- and reverting to dimensional quantities, the force is given in our model by
\begin{equation}
f(t) = 2\pi\mathrm{l}_{\rm g}\,(\lambda L_0 \kappa^2/\mu)^{1/3}\,t^{-1/3}\,.
\label{eq:ForcePowerLaw}
\end{equation}
Here, $\mathrm{l}_{\rm g}$ is a constant of order one depending logarithmically (i.e., weakly) on $\kappa$, $E$, $\mu$, $L_0$, $\lambda$, and $t$. Importantly, while the pre-factor in Eq.~\ref{eq:ForcePowerLaw} depends on the material properties of the plasma membrane and its surroundings and on the experimental conditions, the decay exponent is expected to be universal and equal to $1/3$. 

\begin{figure*}[t]
	\centering
	\includegraphics[width=.9\textwidth]{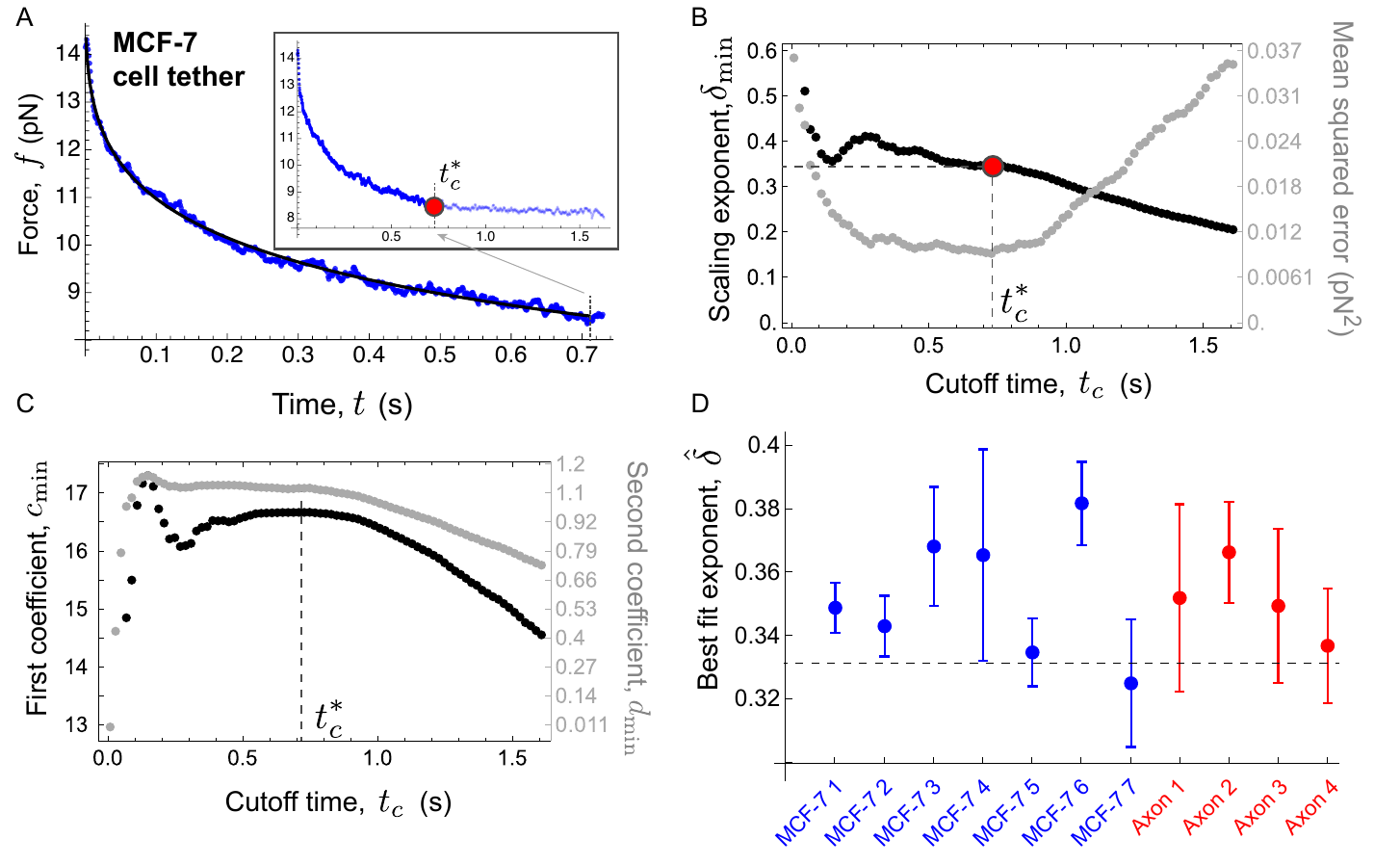}
	\caption{{\bf Experimental force data on suddenly extended tethers confirms prediction of scaling exponent, $\delta=1/3$.} A: Same force data as in Fig.~\ref{fig:Fig1}B, now fitted with the formula $f(t)= c/(d+t^{\delta_{\rm exp}})$. The fit is done from $t=0$ up to some cutoff time $t_c^*=N_c\Delta t_{\rm exp}$, as indicated by red point in the inset.    B:  The cutoff is determined by minimization of the mean squared error, $\chi^2$, as shown in gray. The value of $\delta_{\rm exp}$ is obtained from the least squares fit up to $t_c^*$.
		C:  The best fit values of $c$ and $d$ as a function of $t_c$.  Near $t_c^*$ these are roughly constant.  D:  Repeating the fit procedure for ten other data sets shows very close agreement between the model prediction and experimental tether pulling.  Data sets are courtesy of Pradhan et al.~\cite{Pradhan:2022tf} and Datar et al.~\cite{Datar:2015vp}.}
	\label{fig:Fig4}
\end{figure*}

To test this prediction we re-analysed tether force data from the literature.  We considered data sets of tether pulling from the plasma membrane of two cells types, MCF-7 breast cancer cells~\cite{Pradhan:2022tf} and chick dorsal root ganglion (DRG) neurons~\cite{Datar:2015vp}.  These cell types have distinctly different cortical architectures~\cite{xu2013actin,calzado2016effect}, which one might think would lead to different tether extension responses.  We determined the experimental value of the exponent $\delta$ using a custom fitting procedure that we developed, as described here.  An example of the result of the fit is given in Fig.~\ref{fig:Fig4}A.
To obtain this fit, and noting that since the experimental tether extensions, $\lambda$, were finite, we used the interpolation formula
\begin{equation}
f_{\rm int}(t) = \frac{c}{d+t^\delta}\,,
\label{eq:ForceInterpolation}
\end{equation}
where $c$ and $d$ are constants. 
This expression captures the finite $t\to 0$ behavior, as found experimentally, and the expected power law at longer times.  It also yields fast short time force decay, as the slope of $f_{\rm int}$ diverges as $t$ and $d$ tend to zero.   Equation~\ref{eq:ForceInterpolation} does not, however, capture the very long time limit, in which the force is expected to tend to its unperturbed value set by $\sigma_0$.  This behavior, occurring beyond the time window $\Delta t_{\rm ss}$, is not described by our asymptotic calculation, as alluded to earlier.

To handle this problem we considered the time, $t_c$, over which the fit was performed 
as a free parameter. We did this using the mean-squared error $\chi^2(\delta,c,d,t_c) = \frac{1}{N_c}\sum_{i=1}^{N_c} (f_{\rm int}(t_i)-f_i)^2$.  In this formula,
$t_i$ is the $i$th time point in the data series and $f_i$ is the experimental force at that time.
Also, $N_c$ is the number of data points in the fit, and is related to $t_c$ by 
$t_c=N_c\Delta t_{\rm exp}$, where $\Delta t_{\rm exp}$ the time interval between successive data points.
In the first step of the fit, we minimized $\chi^2$ 
 with respect to its first three arguments, namely, $\delta$, $c$, and $d$.  
The resulting minimizers, denoted with subscript ``min'', are thus functions of $t_c$.  In the second step we minimized $\chi^2(\delta_{\rm min}(t_c), c_{\rm min}(t_c),d_{\rm min}(t_c),t_c)$ with respect to $t_c$ to obtain $t_c^*$ and the best fit parameters.   

An example of the results of the fitting are given in Figs.~\ref{fig:Fig4}B and C, corresponding to the data from Fig.~\ref{fig:Fig4}A. The best fit value of the force decay exponent, $\hat{\delta}\equiv\delta_{\rm min}(t_c^*)$, was found to be $\hat{\delta}\approx 0.34$. It corresponds to a minimum in the mean-squared error.  In Fig.~\ref{fig:Fig4}B we observe that with increasing $t_c$ the exponent $\delta_{\rm min}$ first varies rapidly, then levels off near $t_c^*$, and finally decreases.  Similar behavior is observed for the other parameters, $c_{\rm min}$ and $d_{\rm min}$; see Fig.~\ref{fig:Fig4}C.   These trends align with the validity limits of the asymptotic calculation: for $t_c$ too small, the expected departure from self-similarity leads to poor parameter estimation; for $t_c$ near $t_c^*$ the fit is only weakly sensitive to the number of data points, which makes sense if $\Delta t_{\rm ss}$ is sufficiently large; whereas for $t_c$ too large, the fit becomes poor again.  This is not surprising, since the interpolation formula is invalid for very long times.
The fitting procedure was repeated for 
six more tether data sets from MCF-7 cells (see SI). Figure~\ref{fig:Fig4}D shows that in each case $\hat{\delta}$ is within 10\% of the theoretical value of 1/3.  

We then tested our prediction of material independence of the force power law exponent by re-analyzing data on suddenly extended tethers from the axons of Chick DRG neurons~\cite{Datar:2015vp}. Axons contain a cortex consisting of periodically spaced actin rings that are sequentially crosslinked by spectrin tetramers~\cite{xu2013actin}.
By fitting our model to four data sets~(see SI), we found that the best fit value of the exponent, $\hat{\delta}$, was in each case within a few percent of 1/3 (Fig.~\ref{fig:Fig4}).  
Remarkably, we find that, overall, the best fit exponent $\hat{\delta}$ for each of the eleven sets is within no more than 10\% of the expected value of $1/3\simeq 0.33$, with a sample-wide mean of $\langle \hat{\delta}\rangle = 0.35$ and standard deviation of 0.02.  We note that $\langle\hat{\delta}\rangle \gtrsim 1/3$ is not unexpected and is a slight artefact of the interpolation formula.  Indeed, fitting $f_{\rm int}(t)$ to numerically generated force versus time data yields a value of $\hat{\delta}$ slightly in excess of 1/3 (see SI). 
Taken together, by fitting suddenly extended tether force versus time  from two cell types with our model shows that the decay follows a power law with a universal exponent equal to 1/3. This results from the underlying scale-invariance of the membrane tension relaxation and implies long range membrane flows over scales greater than those set by membrane-adjacent structures. 

%

%
\section{Discussion}
Pulling out membrane tethers from cells, 
dating to the early nineteen seventies~\cite{Hochmuth:1973wa}, is a well established
method to probe membrane viscoelasticity and its coupling to different cell functions.  A directly relevant early example involved suddenly extending tethers  from axons to detect the growth cone-to-soma directed membrane flows in neurons.  This was an important discovery, demonstrating a causal link between membrane tension gradients and spatially extended lipid flows~\cite{Dai:1995wr}.
Tether pulling has also been used to uncover mechanical interaction between the plasma membrane and its surroundings. By comparing pulling rate-dependent tether forces from cells~\cite{Dai:1995tw,Shao:1996wy,Hochmuth:1996tu} and, more recently, protein-decorated vesicles ~\cite{Guevorkian:2015vs,Campillo:2013ur} with bare vesicles~\cite{EVANS199439,Simunovic:2017tj}, where forces are rate-independent, it has been possible to identify the frictional interaction between lipid membrane flows and the membranes surroundings, e.g., a cortex, peri-cellular matrix~\cite{Ghisleni:2024wv}, or some other substrate, as a key parameter affecting membrane tension propagation in cells~\cite{Shi:2018tv,Gomis-Perez:2022vx,Shi:2022uj,De-Belly:2021we,Garcia-Arcos:2024tf}.

Given the age and ubiquity of the technique, it is surprising the theory of tether pulling remains underdeveloped.
Until now there has been, with one exception, no solid theory that relates pulling rate-dependent tether force to spatially extended lipid flows in the surrounding membrane. The one exception considered constant speed tether pulling on cells and the resulting steady-state force~\cite{Brochard-Wyart:2006wv}. 
While various complicated time-dependent pulling protocols have been used to probe membrane tension propagation~\cite{Shi:2018tv,Shi:2022uj,Gomis-Perez:2022vx,Dharan:2025tx}, the simplest, non-trivial pulling assay to test time-dependent membrane response is to very quickly extend  a pre-existing tether on the cell plasma membrane to a new length and measure the force relaxation~\cite{Heinrich:2005vs,Jauffred:2007uj,Li:2002vn,Datar:2015vp,Li:21Optics,Campillo:2013ur,Gomis-Perez:2022vx}. This is superficially similar to a step-strain experiment in rheology, for which the simplest viscoelastic behavior is the Maxwell model (spring in series with a viscous dasphot), which gives a simple exponential decay of the shear stress~\cite{rubinstein2003polymer}.  Perhaps because of this resemblance, attempts at rationalizing step-length membrane tether response have involved fitting a single exponential to the force decay data~\cite{Li:2002vn,Jauffred:2007uj,Khatibzadeh:2013uw}. Others have recognized that this does not account for the fast decay observed at early times, and have opted for a double exponential fit~\cite{Gomis-Perez:2022vx,Pradhan:2022tf}.  It is important to stress that there is no physical reason to model suddenly extended tether forces with a single or double exponential function.  

A different approach to modelling dynamic tether force decay was to assume that sudden extension depletes the tether of mobile proteins and the decaying force reflects a slow influx and diffusion from the membrane to the tether~\cite{Datar:2015vp}.  In other words, the tension gradient derives from a two-dimensional osmotic imbalance of mobile proteins~\cite{ZHANG2024102377}.  The resulting force relaxation is non-exponential, and at long times and for sufficiently large extensions, it can be shown to vary as $t^{-1/4}$. Because of the closeness to our scaling prediction, $f\propto t^{-1/3}$, this model appears to do a good job in fitting tether data pulled from different cell types~\cite{Datar:2015vp,Paraschiv:2021ub}.  Its shortcoming is to assume the tension in the membrane surrounding the tether equilibrates instantaneously, which is difficult to justify in light of double tether experiments revealing slow tension propagation on cortex-coupled plasma membranes~\cite{Shi:2018tv, Gomis-Perez:2022vx,Shi:2022uj,De-Belly:2023vo}.

Interestingly, the first (to our knowledge) published step-length experiments on tethers recognized that their viscoelastic behavior could not simply be Maxwellian~\cite{Heinrich:2005vs}.  Noting the force on a tether pulled at constant speed, $U$, increased sub-linearly with $U$, Heinrich et al. proposed a purely phenomenological ``shear-thinning'' model, which, applied to the case of sudden extension,  predicted a quasi power law decay with a best-fit exponent $\approx 0.25$.  This is remarkably close to our analytical prediction, $f\propto t^{1/3}$.  It also prompted Brochard-Wyart et al.~\cite{Brochard-Wyart:2006wv} to (correctly) recognize there is no need to invoke an alleged shear-thinning to explain the behavior. Using known membrane tube mechanics and mass conservation she arrived at the formula $f\propto U^{1/3}$ in steady-state. This relation reflects the membrane tension gradient needed to balance frictional resistance and power the lipid influx into the extending  tether, and not coincidentally it is the same exponent as we find here for sudden tether extension. 

The model of Ref.~\cite{Brochard-Wyart:2006wv} has become the gold standard for quantitatively interpreting the link between membrane-cortex friction and tension gradients~\cite{Diz-Munoz:2010wc,Diz-Munoz:2013we,Simunovic:2017tj,SITARSKA202011,Bergert:2021wq}.  Yet, with our current model we can identify a conceptual problem in this work. Because Ref.~\cite{Brochard-Wyart:2006wv} considered steady-state incompressible, two-dimensional membrane flows induced by tether pulling, they implicitly assumed instantaneous tension propagation, which cannot be correct. This assumption leads to a diverging tension at long distances from the tether --- related to Stokes' paradox in fluid mechanics ---, and to circumvent it they were forced to introduce an unphysical long-distance cutoff $R$. With the insight of our model, this issue can be avoided by simply
replacing $R$ with the length scale $\sqrt{D t}$ over which tension has propagated a time $t$ after a perturbation and over which the model membrane behaves as an incompressible, two-dimensional fluid.  

The connection of our work to Ref.~\cite{Brochard-Wyart:2006wv} raises the question, what can we learn from sudden tether pulling that we cannot from working at constant speed? There are 
folded membrane structures that interact bi-directionally with membrane tension~\cite{Sinha:2011ur,Lolo:2023ve}. They also change depending on membrane-to-cortex attachment state~\cite{Itoh:2023aa}, membrane composition~\cite{10.7554/eLife.55038} and osmotic stress~\cite{Roffay:2021tw}.  These  changes, which may involve membrane-subjacent cortical dynamics~\cite{De-Belly:2023vo,lafoya2023consumption} and/or endo- and exo-cytosis~\cite{alonso2024cell}, occur on seconds to minutes time scales.  How such a highly dynamic scenario might modify membrane tension would be more readily probed by single or repeated sudden tether extension than by 
constant speed pulling, which only reports on long time, steady-state changes.

To conclude, our model makes a specific prediction about the scaling of the tether force, decoupling a universal exponent describing decay from a material-dependent prefactor (Eq.~\ref{eq:ForcePowerLaw}). We have tested and confirmed our model on two cell types, for which tethers were pulled from the axonal membrane of neurons~\cite{Datar:2015vp} and from the plasma membrane of MCF-7 breast cancer cells~\cite{Pradhan:2022tf}.  These two cases involve distinct cortical architecture and hence different local membrane-cortex interactions~\cite{xu2013actin,Padilla-Rodriguez:2018tl}.  The robust scaling of the force, $f\sim t^{1/3}$, is rooted in the self-similar structure of membrane tension diffusion, which is an attractor solution for a range of experimentally-realizable problems with large tether extension factors. It suggests also that long range membrane flows, independent of local details, relax the tether force.  It would be interesting to experimentally challenge our prediction by performing sudden tether extension on different parts of the plasma membrane, such as adherent versus non-adherent.  While the exponent should be robust, a careful mapping of the prefactor could inform on spatial inhomogeneity in membrane-to-substrate attachment.  
Relaxation of membrane tension involving mechanisms not involving lipid flows could potentially alter our picture of tension relaxation~\cite{Rangamani:2022ti}. For example, endo- and exo-cytosis and conformation changes of membrane folded structures are known to be regulated by, and in turn can relax, tension perturbations~\cite{Dai:1995un,Raucher:1999vp,morris2001cell,Sinha:2011ur,Gauthier:2011wc,Roffay:2021tw,Lolo:2023ve}. Sudden tether extension could provide quantitative signatures of these processes.  


\vspace{.5cm}
%
\pagebreak
\section{Materials and Methods}
\begin{center}
	\bf{Non-dimensionalization of boundary value problem}
\end{center}
The diffusion problem, Eqs.~\ref{eq:DiffusionEquation}-\ref{eq:BCs}, was solved numerically after first making the boundary value problem dimensionless.  
Lengths are expressed in units of $r_0$ (the tether radius prior to sudden extension), times in $r_0^2/D$, and tension in $\sigma_0$ (tension prior to extension).  Furthermore, assuming axi-symmetry with respect to the tether axis, i.e., $\sigma=\sigma(r,t)$, it follows that the diffusion equation, the boundary condition far from the tether, and the two conditions at $r=r_{\rm t}$ (Eqs.~\ref{eq:TetherRadius} and \ref{eq:BCflux}) may be re-written as
\begin{eqnarray}
	&\partial_t\sigma = \frac{1}{r}\partial_r (r\partial_r \sigma)\,, 
	&\quad\sigma(r\to\infty,t)=1\,; \label{eq:PDE}\\[.2cm]
	&\left.\partial_r \sigma\right|_{r=r_{\rm t}} = -{\displaystyle\frac{2\lambda}{b} \frac{d r_{\rm t}}{dt}} \,,
	& \quad r_{\rm t}= \frac{1}{\sqrt{\sigma(r=r_{\rm t},t)}}\,, \label{eq:BCs}
\end{eqnarray}
for $r\in [r_{\rm t},\infty)$.
The initial condition is given by
\begin{equation}
	\sigma(r,0)=
	\begin{cases}
		\lambda^2\,, & r=r_{\rm t }(0)\,; \\
		1\,, & r>r_{\rm t }(0)\,. 
	\end{cases}
	\label{eq:ICs}
\end{equation}
Equations~\ref{eq:PDE}-\ref{eq:ICs} define a free boundary value problem, since the tether position $r_{\rm t}(t)$ is unknown and must be determined as part of the solution.  

The two dimensionless parameters in the problem are the extension ratio, $\lambda=L/L_0$, and $b=\sqrt{2\kappa\sigma_0}/EL_0$ (Eq.~\ref{eq:bDef}).
The latter one can be estimated as follows.
Taking $\kappa=2\times 10^{-19}$ ~\cite{Shi:2018tv};
$\sigma_0=(1-10)\times 10^{-5}$ N/m, $E=	(4-200)\times 10^{-4}$ N/m~\cite{Gomis-Perez:2022vx}; and $L_0=5$ $\upmu$m~\cite{Datar:2015vp}; it follows that $b\approx 10^{-5}-10^{-3}$.

\begin{center}
	\bf{Numerical resolution method}
\end{center}

To numerically solve the moving boundary value problem in the form Eqs.~\ref{eq:PDE}-\ref{eq:ICs} we used the methods outlined in~Refs.~\cite{Crank1957__TwoMethodsNumerical,crank1987free}.
When trying to solve a moving boundary  value problem on a fixed grid, the moving boundary generally falls between grid points. To overcome this problem, special methods are needed to deform the grid, to ensure the free boundary always falls on the grid, or to obtain special formulas allowing for unequal spacing around the moving boundary. 
Here, we adopt the latter strategy following ~\cite{Crank1957__TwoMethodsNumerical, crank1987free}.

We first define 
\begin{equation}
\sigma_{i,j} = \sigma(i\Delta t,j\Delta r),\quad\text{for}~i \in [0,n_t]~\text{and}~j \in [0,n_r]\,,
\end{equation}
\noindent where $n_t$ ($n_r$) defines the width of the discretized time (space) domain in units of the time step $\Delta t$ (grid spacing $\Delta r$). We then call $j_{\rm t}(t)$ the closest grid point before the moving boundary such that $j_{\rm t}(t)\Delta r< r_{\rm t}(t)\leq(j_{\rm t}(t)+1)\Delta r$, and we define $p\in]0,1[$ such that $r_{\rm t}(t)=(j_{\rm t}(t)+p(t))\Delta r$ to locate $r_{\rm t}(t)$ with respect to the grid.

Away from the free boundary, i.e., for $j$ such that $j_{\rm t}+1<j<n_r$, we discretize the PDE for membrane tension given by Eq.~\ref{eq:PDE} using a forward difference in time and a central difference in space such that:
\begin{equation}
\label{eq:PDEfinitedifference}
\frac{\sigma_{i+1,j}-\sigma_{i,j}}{\Delta t}=\frac{\sigma_{i,j+1}-2\sigma_{i,j}+\sigma_{i,j-1}}{\Delta r^2}+\frac{1}{j\Delta r}\frac{\sigma_{i,j+1}-\sigma_{i,j-1}}{2\Delta r}\,.
\end{equation}
However, this equation does not hold near the $r_{\rm t}$ boundary, i.e., for $j=j_{\rm t}+1$, as the equation is not defined for $r=j_{\rm t}\Delta r$. To enforce the boundary condition defined for $r_{\rm t}$ despite it not being placed on the grid, we built special formulas based on a Lagrangian interpolation.

Following the steps in Refs. \cite{Crank1957__TwoMethodsNumerical}, the three-point Lagrangian interpolation formula for the tension $\sigma$ is defined by
\begin{equation}
\label{eq:sigmainterpolation}
\sigma(r)=\sum_{n=0}^2l_n(r)\sigma(a_n)\,,
\end{equation}
\noindent where
\begin{equation}
l_n(r)= \frac{p_2(r)}{(r-a_n)p_2'(a_r)},\quad p_2(r)=(r-a_0)(r-a_1)(r-a_2)\,,
\end{equation}
\noindent and $p_2'(a_n)$ is the derivative of $p_2(r)$ with respect to $r$ evaluated at $r=a_n$.

Taking 
\begin{equation}
a_0=(j_{\rm t}+p)\Delta r,\quad a_1=(j_{\rm t}+1)\Delta r,\quad a_2=(j_{\rm t}+2)\Delta r\,,
\end{equation}
\noindent and deriving Eq. \ref{eq:sigmainterpolation} twice, one then obtains for $j=j_{\rm t}+1$:
\begin{align}
\frac{\sigma_{i+1,j_{\rm t}+1}-\sigma_{i,j_{\rm t}+1}}{\Delta t}&= \nonumber \\
&\hspace{-2cm}	\frac{2}{\Delta r^2} \left(\frac{\sigma_{i,r_{\rm t}}}{(1-p)(2-p)} - \frac{\sigma_{i,j_{\rm t}+1}}{1-p} + \frac{\sigma_{i,j_{\rm t}+2}}{2-p}\right)\nonumber \\ 
&\hspace{-3cm}- \frac{1}{(j_{\rm t}+1)\Delta r^2}\left(\frac{\sigma_{i,r_{\rm t}}}{(1-p)(2-p)}-\frac{p\sigma_{i,j_{\rm t}+1}}{1-p}-\frac{(1-p)\sigma_{i,j_{\rm t}+2}}{2-p}\right)\,,
\label{eq:sig_jtube}
\end{align}
\noindent where $\sigma_{i,r_{\rm t}}$ is given by the second member in Eq.~\ref{eq:BCs}. 

Finally, using the same interpolation procedure, the position of $r_{\rm t}$ is given by:
\begin{align}
\frac{r_{\rm t}((i+1)\Delta t)-r_{\rm t}(i\Delta t)}{\Delta t}&= 
-\frac{b\, r_{\rm t}(i\Delta t)}{2\lambda\Delta r}\left(\frac{(2p-3)\sigma_{i,r_{\rm t}}}{(1-p)(2-p)} 
\right. \nonumber \\
&\hspace{-1cm} \left.
+ \frac{(2-p)\sigma_{i,j_{\rm t}+1}}{1-p} + \frac{(p-1)\sigma_{i,j_{\rm t}+2}}{2-p} \right)\,.
\label{eq:rtetherfinitedifference}
\end{align}
To summarize, the steps to obtain the numerical solutions are:
\begin{itemize}
	\item At $t=0$, knowing $\sigma(r,t=0)$ for all values of $r$ from Eq. \ref{eq:ICs}, deduce $r_{\rm t}(t=0)$ using Eq. \ref{eq:BCs} and the corresponding values of $j_{\rm t}(t=0)$ and $p(t=0)$ satisfying the definition given above.
	\item For $t>0$, iterate over the following steps:
	\begin{itemize}
		\item For $n_r>j>j_{\rm t}+1$, compute $\sigma_{i,j}$ from Eq. \ref{eq:PDEfinitedifference}.
		\item For $j=j_{\rm t}+1$, compute $\sigma_{i,j_{\rm t}+1}$ from Eq. \ref{eq:sig_jtube}.
		\item For $j=n_r$, $\sigma_{i,n_r}$ is given by the second member in Eq.~\ref{eq:PDE}.
		\item For $j\leq j_{\rm t}$, we take $\sigma_{i,j}=\sigma_{i,r_{\rm t}}$.
		\item Update the value of $r_{\rm t}$ using Eq.~\ref{eq:rtetherfinitedifference}.
		\item Update the values of $j_{\rm t}$ and $p$.
	\end{itemize}
\end{itemize}

\vspace{.5cm}
\begin{center}
	\bf{Parameter values used in Figs.~\ref{fig:Fig2} and \ref{fig:Fig3}}
\end{center}
For Fig.~\ref{fig:Fig2}A the following number of time and space steps, as defined in the Numerical Resolution Method, were used: $n_{t}=10^7$ and $n_r=7000$.  The temporal and spatial step sizes  were $\Delta t=10^{-5}$ and $\Delta r = 500/7000\approx 0.71428$.  The extension factor $\lambda$ and parameter $b$ were chosen to be $\lambda=128$ and $b=0.1$.  The choice of $b$, while still less than one, was chosen to be orders of magnitude larger than experiment (see Non-dimensionalization of boundary value problem) to avoid the excessive computing time that would be needed to observe a significant force decay had we chosen a smaller $b$. 

For Fig.~\ref{fig:Fig2}B the number of time steps $n_t$ and the parameter $b$ were the same as in A. The step size $\Delta t$ was chosen to be $\Delta t=10^{-5}$.  For $\lambda \in \{2,4,8,16\}$ we used $n_r=3\times 10^4$ and $\Delta r=1/30$.  
For $\lambda \in \{32, 64, 128\}$ we used $n_r=2\times 10^4$ and $\Delta r =1/20 = 0.05$. 

For Fig.~\ref{fig:Fig3}A we used the same value of $b$ as in Fig.~\ref{fig:Fig2}, $n_t=2\times 10^6$, and $\Delta t=10^{-5} $.
 For $\lambda \in \{2,4,8,16,32\}$ we used $n_r=10^4$ and $\Delta r=0.15$.  For $\lambda \in \{64,128\}$ we used $n_r=2\times 10^4$ and $\Delta r=0.075$.  

Finally, for Fig~\ref{fig:Fig3}B we used the values for $b$ and $\lambda$ as in Fig.~\ref{fig:Fig2}A.  We chose $n_t=3\times 10^6$, $\Delta t= 1/3\times 10^{-4}$, $n_t=3\times 10^4$, and $\Delta r=15/300$.  

\vspace{.5cm}
\begin{center}
	\bf{Determination of scaling exponents $\beta$, $\alpha$, and $\delta$}
\end{center}
In general, an analytical solution to the non-linear eigenvalue problem Eqs.~\ref{eq:PDE}-\ref{eq:ICs} is not available. 
However, if we assume that the extension factor $\lambda\gg 1$ and $b\ll 1$ we expect there to be an intermediate time scale over which the solution 
is independent of initial and final conditions.   In this regime we will find scaling 
and self-similar behavior in the tether radius and tension profile. This approach borrows ideas from the theory of Intermediate Asymptotics~\cite{barenblatt_1996}.  In carrying out 
the asymptotic calculation of $\sigma(r,t)$ we will find the exponents, $\alpha$, $\delta$, and $\gamma$, as 
well as the function $g(u)$ and the coefficent $a$.

To begin, we propose the following Ansatz for the tension:
\begin{equation}
\sigma(r,t) = t^{-\alpha} g\left(\frac{r}{t^\beta}\right)\,.
\label{appendeq:TensionAnsatz}
\end{equation}
Substituting this into the PDE, Eq.~\ref{eq:PDE}, we obtain
\begin{align}
	g''(u) +\left(\frac{1}{u}+\beta\,u\,t^{2\beta-1}\right)g'(u)+\alpha\,t^{2\beta-1}g(u)&=0\,,
	\label{appendeq:DiffusionEquationforu}
\end{align}
where the scaling variable $u\equiv r/t^\beta$ and the primes denote differentiation with respect to $u$.  
Requiring that all three terms above are balanced for $u\to 0$ and $u\to \infty$, we 
find that 
\begin{equation}
\beta=1/2\,,
\label{appendeq:beta}
\end{equation}
as expected for a diffusion problem. 

Next, we assume that the tether radius has the scaling form 
\begin{equation}
r_{\rm t}= a\,t^{\delta}\,,
\label{appendeq:TetherRadius}
\end{equation}
where $a>0$ and $\delta>0$ are constants.  We assume that $\delta < 1/2$, which is confirmed below. 
Substituting the self-similar expression Eq.~\ref{appendeq:TensionAnsatz} into the boundary conditions, Eq.~\ref{eq:BCs}, and replacing $r_{\rm t}$ with Eq.~\ref{appendeq:TetherRadius}, we obtain
\begin{align}
t^{\delta-\alpha/2}&=\frac{1}{a\sqrt{g(a\,t^{\delta-1/2})}}
\label{appendeq:scalingBC1}
\\
t^{1/2-\alpha} g'(a\,t^{\delta-1/2})&=-\frac{2\delta\lambda}{b}\,.
\label{appendeq:scalingBC2}
\end{align}
The exponents $\alpha$ and $\delta$ are then determined as follows.  At long times the re-scaled position of the tether radius $u_{\rm t}\equiv a\,t^{\delta-1/2}$ is $\ll 1$.  In this regime it is readily seen from Eq.~\ref{appendeq:DiffusionEquationforu} that
\begin{equation}
g(u)_{u\ll 1}= c_1 \log{(u)}+c_2\,,
\label{appendeq:gsmallu}
\end{equation}
where $c_1$ and $c_2$ are integration constants.
Note that the above expression corresponds to the incompressible membrane limit, since the radial flow is $v\propto \partial_u \sigma \propto 1/u$, which is divergence-less. Substituting Eq.~\ref{appendeq:gsmallu} into Eqs.~\ref{appendeq:scalingBC1}-\ref{appendeq:scalingBC2}, it follows that 
\begin{align}
 c_1 t^{1-\alpha-\delta}&=-\frac{2 a\,\delta\lambda}{b}		\label{appendeq:scalingBC3}\\
	t^{\delta-\alpha/2}&=\frac{1}{a\sqrt{c_1\log{(a t^{\delta-1/2})}+c_2}}\,.	
\label{appendeq:scalingBC4}
\end{align}
From the first equation above we find that 
\begin{equation}
\alpha+\delta=1
\label{appendeq:relationalphadelta}\,;
\end{equation}
and from the second equation, up to a logarithmic correction, we find $\delta=\alpha/2$ and thus 
\begin{equation}
\delta=1/3\:,\quad \alpha = 2/3\,.
\label{appendeq:deltaalphavalues}
\end{equation}
Therefore, we have determined the exponents governing the growth of the tether radius, namely $r_{\rm t}\propto t^{1/3}$ (see Fig.~\ref{fig:Fig3}a),
and the decay of the maximum in tension, namely $\sigma_{\rm max}\propto t^{-2/3}$ (see Fig.~\ref{fig:Fig2}b). 

\vspace{.5cm}
\begin{center}
	\bf{Determination of the function $g(u)$ and the pre-factor $a$}
\end{center}
Knowing $\beta$, $\alpha$, and $\delta$, with a bit of work the quantities $g(u)$ and $a$ can be found.  Substituting $\beta=1/2$ and $\alpha=2/3$ into Eq.~\ref{appendeq:DiffusionEquationforu}, and demanding that for large $u$ it does not blow up, we find it is equal to a Meijer-G function~\cite{Mathematica}:
\begin{equation}
g(u) = A\,G_{1,2}^{2,0}\left(\frac{u^2}{4}\Big|
\begin{array}{c}
\frac{1}{3} \\
0,0 \\
\end{array}
\right)\,,
\label{appendeq:gofuMeijerG}
\end{equation}
where $A$ is a constant to be determined.
The asymptotic expansion of this solution for small $u$
is 
\begin{equation}
g(u)\sim - \frac{2A}{\Gamma(1/3)}\left[\log (u)+\gamma -\log (2)+\frac{1}{2}\psi(1/3)\right]\,,
\label{appendeq:gasymptotic}
\end{equation}
where $\Gamma(z)$ is the Gamma-function, $\gamma\simeq 0.577$ is Euler's constant, and $\psi(z)$ is the digamma function, with $\psi(1/3)\simeq -3.13$. Comparing Eqs.~\ref{appendeq:gsmallu}
and \ref{appendeq:gasymptotic}, we see that 
\begin{equation}
c_1 = - \frac{2A}{\Gamma(1/3)}
\end{equation}
and
\begin{equation}
\frac{c_2}{c_1}=\gamma-\log(2)+\frac{1}{2}\psi(1/3)=:\tilde{\gamma}\simeq -1.682\,.
\end{equation}
As a result, from Eqs.~\ref{appendeq:scalingBC3}-\ref{appendeq:scalingBC4} 
we obtain
\begin{equation}
a^3\log{\left(a e^{\tilde{\gamma}} t^{-1/6}\right)^{3}}=-\frac{9b}{2\lambda}\,.
\label{appendeq:ImplicitEquationFora}
\end{equation}
An approximate solution to the above equation for $a$ is found by
first letting $w=\frac{1}{3}\log{(a e^{\tilde{\gamma}} t^{-1/6})}$.  
Then, the equation to solve becomes $w\,e^w = -9b/2\lambda\,e^{3\tilde{\gamma}} t^{-1/2}$. 
Finally, solving $w\,e^w=-z$ for $z>0$ and $z\ll 1$ yields $w\simeq \log(z)-\log(-\log(z))$.   Thus, 
\begin{equation}
a\simeq \left(\frac{9 b}{2\lambda}\right)^{1/3}\left[-\log{(9b e^{3\tilde{\gamma}} t^{-1/2}/2\lambda)}\right]^{-1/3}\,.
\label{appendeq:FinalEquationfora}
\end{equation}
Furthermore, the coefficient $A$ in Eq.~\ref{appendeq:gofuMeijerG} is
\begin{equation}
A\simeq \frac{\Gamma(1/3)\lambda}{3b} \left(\frac{9 b}{2\lambda}\right)^{1/3}\left[-\log{(9b e^{3\tilde{\gamma}} t^{-1/2}/2\lambda)}\right]^{-1/3}\,.
\label{appendeq:expressionforA}
\end{equation}
The expression for $A$ given in this equation along with Eq.~\ref{appendeq:expressionforA} give an analytical expression for $\sigma(r,t)$ for intermediate time scales (Fig.~\ref{fig:Fig3}A).

\vspace{.5cm}
\begin{center}
	\bf{Determination of the logarithmic factor $\mathrm{l_g}$}
\end{center}
Finally, with knowledge of $a$ (Eq.~\ref{appendeq:FinalEquationfora}), the tether force is determined.  Reverting to dimensional units, we obtain $f= 2\pi\mathrm{l}_{\rm g}\,(\lambda L_0 \kappa^2/\mu)^{1/3}\,t^{-1/3}$  (Eq.~\ref{eq:ForcePowerLaw}), where 
\begin{equation}
\mathrm{l_g}\equiv \left[-\frac{2}{9}\log{\left(
	\frac{9 
		\kappa  e^{3\tilde{\gamma}}}{2\lambda L_0 \mu^{1/2} E^{3/2} t^{1/2}} 
	\right)}\right]^{1/3}\,.
\vspace{0.2cm}
\end{equation}
We note that $\mathrm{l_g}$, while time-dependent, can be seen to be slowly varying. Taking $\lambda=1-10$, the parameter values from Non-dimensionalization of boundary value problem, $\mu^{-1}=1.7\times 10^9$ Pa.s/m~\cite{Shi:2018tv}, and $t=1-10$ s, we find that $\rm{l_g}\approx 2$. 
%
%
\vspace{1cm}

\begin{acknowledgments}
	We wish to thank Julien Browaeys for useful discussions about data analysis.  We are also grateful to Patricia Bassereau, Jean-Baptiste Manneville, Shivani Dharmadhikari, and Jean-L\'eon Ma\^itre for their careful reading and comments on the manuscript.
\end{acknowledgments}

\section*{Author contributions}
EL and ACJ designed and performed research and wrote the paper.  EL developed the code for numerical solution and ACJ developed the analytical model and performed the data re-analysis. 

\vspace{.5cm}

\bibliography{References}

\end{document}